\documentstyle[12pt]{article}

\def\sqr#1#2#3#4{{\vcenter{\vskip -#3 pt\hbox{\kern #4 pt\vbox{\hrule 
height
.#2 pt\hbox{\vrule width .#2 pt height #1 pt\kern #1 pt\vrule width 
.#2 pt}
\hrule height .#2 pt}\kern #4 pt}}}}

\def\QED{\hbox{\kern 1pt\vrule width 3pt height 7pt}}
\def\p{\partial}

\def\d{\delta}

\def\a{\alpha}
\def\b{\beta}
\def\g{\gamma} 
\def\om{\Omega} 

\def\del{\delta}

\def\be{\begin{equation}}
\def\bea{\begin{eqnarray}}

\def\ee{\end{equation}}
\def\eea{\end{eqnarray}}

\def\C{\rm {I\kern-.520em C}}
\def\R{\rm {I\kern-.150em R}}
\def\w{\rm {WZNW }}

\def\D{\partial}
\def\la{\lambda}

\begin{document}

\begin{titlepage}
\begin{flushleft}
       \hfill                      IPM-95-119\\
       \hfill                      hep-th/9601068\\ 
\end{flushleft}
\vspace*{3mm}
\begin{center}
{\LARGE WZNW Models from Non-Standard Bilinear Forms\\}

\vspace*{12mm}
{\large H. Arfaei, \,\, S. Parvizi } \\

{\it Institute for Studies in Theoretical Physics and Mathematics, \\
 P.O.Box 19395-1795, Tehran, Iran } \\
{\it  Department of Physics, Sharif University of Technology, \\
\it  P.O.Box 11365-9161, Tehran, Iran \/}\\
\end{center}
\vspace*{15mm}

\begin{abstract}

We study the WZNW models based on nonstandard bilinear forms. We approach 
the problem from algebraic, perturbative and functional exact methods. It 
is shown that even in the case of integer $k$ we can find irrational CFT's. 
We prove that when the base group is noncompact with nonabelian maximal 
compact subgroup, the Kac-Moody representations are nonunitary. 

\end{abstract}

\end{titlepage}

\section{Introduction:}
\par {Certain WZNW models do not allow the standard formulation based on the
Killing form as the bilinear form in their algebra used to define the action.
These models are based on a group $G$ whose Killing form is 
singular (so $G$ is a non-semisimple lie group). In some 
cases such as centrally extended non-semisimple groups existence of another 
invariant bilinear form \cite{Cangemi} comes to our help, enabling us to 
construct a nonsingular invariant 
bilinear form to be used in the definition of the action \cite{Nappi}. Such 
WZNW models have been used to construct cosmological models. A non-semisimple 
group $E^c_2$ in this manner is used to describe a homogeneous space with 
central charge four.}
\par {These models were approached in another way in \cite{Olive} and 
were generalized to include $E^c_d$ (centrally extended 
Euclidean group in $d$ dimensions) \cite{Sfetsos} and double extension of 
the lie groups \cite{Ofarrill}. Also it was shown that the Sugawara 
construction can be extended to such models with an arbitrary 
invariant nondegenerate bilinear form \cite{Nouri}.} 
\par {In this paper we consider several aspects of these non standard models.
We show that conformal invariance is achieved in a similar way to the 
standard case. In the Sugawara construction of the
Virasoro generators one needs to know the inner product of two currents. 
This is given by a     
bilinear form of the group which is subject to the Affine-Virasoro Master 
Equation (AVME) \cite{Halpern}. We show that certain symmetry requirements 
on the Sugawara construction restrict the solutions of the AVME to a 
unique form in agreement with \cite{Nouri}. From this solution one can 
find the Virasoro's central charge \cite{Nouri}.} 
\par {As a special case we consider $SO(3,1)$ group to exemplify the 
general arguments. $SO(3,1)$ has more than one bilinear form with 
the desired properties. We find central charges corresponding to these 
nonstandard bilinear forms, then look for representations of the 
corresponding Kac-Moody algebras. We also find the conformal weights. 
Since the Kac-Moody level must be an integer for the Lorentz group, 
the standard WZNW models based on it can have only 
rational conformal central charges and weights, in contrast to the present 
case with real central charges and weights which is due to extra parameter in 
the nonstandard bilinear form. This is interesting that despite the fact that 
$k$ must be an integer we can obtain irrational central charges and 
noninteger effective level.}
\par{On the other hand, WZNW models based on non-compact groups have also 
been considered. In \cite{Balog} it is shown that 
$SU(1,1)$ theory is nonunitary, but latter works \cite{Petro,Moham,Hwang} 
showed that string models for this group is indeed unitary thanks to 
Virasoro conditions (no-ghost theorem). There is an argument in \cite{Bars} 
for nonunitarity of the Kac-Moody of $SO(3,1)$ in large limit of the 
level. In this paper we generalize these results to the representations of 
any Kac-Moody algebra based on a non-compact group with nonabelian maximal 
compact subgroup. Whether despite nonunitarity of Kac-Moody representations
unitary Virasoro representations can be extracted shall be addressed 
elsewhere.}
\par {In the second part of this paper we treat the model perturbatively, 
finding the $\b$-function 
and generating functional to all orders of perturbations. A non trivial
question about the path integral is whether path 
integral measure receives any changes due to the change of bilinear 
form, indeed we show in the appendix that the choice of bilinear form in the 
path integral measure is independent of the choice of bilinear form in the 
action. Finally in this approach we calculate the trace anomaly.
We also analyze the model by functional method obtaining the interesting 
result of the renormalization of the bilinear form which corresponds to 
$k$ renormalization in ordinary WZNW models.}


\section{General Properties; Conformal and Affine Structures:}
\par {Let us consider a WZNW action on a Riemann surface $\Sigma$ 
whose topology is $S^2$:
\be  
    S = \frac{1}{4\la^2}\int_{\Sigma}d^2xTr(\D_{\mu}U\D_{\mu}U^{-1})+
   \frac{k}{12\pi}\int_{B}d^3x\epsilon^{\mu\nu\rho}Tr(\D_{\mu}UU^{-1}
       \D_{\nu}UU^{-1}\D_{\rho}UU^{-1}).    
\ee
$B$ is a three-dimensional manifold whose boundary is $\Sigma$,  
and $U$ is a map from $\Sigma$ (or its extension $B$) to the group $G$. 
The path integral should be independent of the extension of $U$ to $B$. 
When $\Pi_3(G)$, the third homotopy group of $G$ is nontrivial, parameter $k$ 
in the action is restricted. On the other hand, if $\Pi_3(G)=0$  there is no 
restriction on $k$ and it can be any real number. The topology of a 
non-compact group $G$ is the same as the topology of $H \times \R^{d_G-d_H}$ 
in which  $H$ is maximal compact subgroup, implying $\Pi_3(G)=\Pi_3(H)$. 
When  $H$ is abelian, $\Pi_3(H)=0$ and $k$ can be any real number, and when 
$\Pi_3(H)={\bf Z}$ it is restricted to be an integer.} 
\par{If we take $U^{-1}\D_{\mu}U = A^{a}_{\mu}T_{a} $ in which $T_{a}$'s 
are generators of the group $G$; the WZNW action can be written as:
\be  \label{nWZNW}
    S = \frac{1}{2\la^2}\int_{\Sigma}d^2x\om_{ab}A^{a}_{\mu}A^{b}_{\mu}   
      + \frac{k}{12\pi}\int_{B}d^3x\epsilon^{\mu\nu\rho}\om_{cd}f^{d}_{ab}
      A^{a}_{\mu}A^{b}_{\nu}A^{c}_{\rho},        
\ee
where $f^{c}_{ab}$ is structural constant and $ \om_{ab} $ is 
Cartan-Killing metric of the group $G$. $\om_{ab}$ is a bilinear form which 
can be constructed from 
\be \label{cvom}
     -c_v\om_{ab}=f^{d}_{ac}f^{c}_{bd},  
\ee
(or if $T$'s are taken in the fundamental representation 
$ \om_{ab} = -2 Tr(T_{a}T_{b})$ ) 
in which $c_v$ is the second Casimir of the group.}
\par {In a more compact notation one can represent the action as:}
\be   \label{ggwz1}
      S = \frac{1}{2\la^2}\int_{\Sigma}d^2x<A_{\mu},A_{\mu}>_{\om}   
	+ \frac{k}{12\pi}\int_{B}d^3x\epsilon^{\mu\nu\rho}
	<[A_{\mu},A_{\nu}],A_{\rho}>_{\om},       
\ee        
where 
$  <x,y>_{\om} = \om_{ij}x^iy^j $  is an inner product of x and y, 
two vectors in the Lie algebra of the group. 
\par {We can generalize the standard WZNW model (\ref{nWZNW}) by 
considering 
a bilinear form $M$ instead of $\om$ which has to have three 
properties: 
Firstly, $M$ is a symmetric bilinear form: 
\be \label{symm}
<x,y>_M = <y,x>_M, 
\ee
which is a property of real inner 
products; secondly, it is an invariant of the group which will guarantee
the affine symmetry of the model; $ <UxU^{-1},UyU^{-1}>_{M}  = <x,y>_{M} $  
(for all $U$ in $G$), or equivalently 
$  <[x,y],z>_{M} + <y,[x,z]>_{M} = 0 $. This condition can be expressed as:
\be   \label{inv}
     f^{d}_{ab}M_{cd}+f^{d}_{ac}M_{bd} = 0,         
\ee
where, $ <T_{i},T_{j}>_{M} = M_{ij} $; 
and finally, it has an inverse $M^{ab}$,i.e.:} 
\be \label{invers}
M^{ab}M_{bc} =\d^{a}_{ c}.
\ee
\par {For several simple and semi-simple groups $M$ has a unique form 
which is just $ \om $ the Cartan-Killing form, however, 
there are some groups with more than 
one bilinear form with the above properties. Example is $SO(3,1)$ or 
SL(2,\C)  group.}
\par {Consider the algebras of $SO(3,1)$:
\be
  [J_a,J_b]=\epsilon_{abc}J_c
\ee  
\be
  [J_a,K_b]=\epsilon_{abc}K_c
\ee
\be
  [K_a,K_b]=-\epsilon_{abc}J_c,
\ee
 $\om$ can be derived to be :
\be
 \om =\pmatrix{{\bf 1}_3 & {\bf 0} \cr 
	       {\bf 0} & {\bf -1}_3 \cr}
\ee
here $ {\bf 1}_3 $ is the  3 $\times$ 3 unit matrix; and $c_v = 4$.  
The first(second) three values of $i$ and $j$ in $\om_{ij}$ are 
corresponding to 
$J$'s ($K$'s). This Killing-form corresponds to the first Casimir of the 
group: 
\be
    J^2-K^2 = \om_{ij}T^{i}T^{j}.
\ee
Another invariant bilinear form is introduced via the other Casimir:
\be
    2J\cdot{K} = \om'_{ij}T^{i}T^{j},
\ee
where} 
\be
      \om' =\pmatrix{{\bf 0} & {\bf 1}_3 \cr 
		     {\bf 1}_3 & {\bf 0} \cr}
\ee      
\par {One could take $M$ to be any linear combination of these two 
bilinear  
forms:
\be \label{mso3}
      M = \om + \a \om',  
\ee
where $\a$ is a constant.}
\par {Furthermore, for a non-semi-simple group the Killing-form is 
degenerate and the 
corresponding WZNW action is ill-defined, nonetheless for the most general 
known cases of double extended groups in \cite{Ofarrill}, it is possible 
to find another bilinear form with required properties (\ref{symm},
\ref{inv},\ref{invers}).
$M$ can be introduced as a linear combination of these
two bilinear forms that is non-degenerate, invariant and symmetric.}
\par {Using $M$ the WZNW action could 
be written as: }
\be   \label{gWZNW}   
      S = 
\frac{1}{2\la^2}\int_{\Sigma}d^2xM_{ab}A^{a}_{\mu}A^{b}_{\mu}   
	+ 
\frac{k}{12\pi}\int_{B}d^3x\epsilon^{\mu\nu\rho}M_{cd}f^{d}_{ab}
	A^{a}_{\mu}A^{b}_{\nu}A^{c}_{\rho}.       
\ee
\par {Here topological nature of WZ term is important. As in the standard
WZNW models (\ref{nWZNW}), we need to find the third homotopy group of  $H$
the maximal compact subgroup, because the change of bilinear form on $G$
can not change its topological properties. The topological term restricts 
the values of $k$ to integers when $\Pi_{3}(G) = \Pi_{3}(H) $  is nontrivial.
For example in the case of $SL(2,\C)$ where $H$ is $SU(2)$ restriction of 
$M$ to $H$ is just $\om$ and hence $k$ is integer, however, there is no 
restriction on $\a$.}
\par {Now we are ready to derive algebras from symmetries of the model.
First, let us introduce conserved currents;}
\be
 J_{+a}=k<T_a,U^{-1}\p_{+} U>_M.
\ee
\par {Following \cite{Witten}, one can find the Kac-Moody algebra 
from the action (\ref{gWZNW}) by calculating Poisson brackets of 
currents. The resulting algebra can be stated as an OPE;}
\be \label{jj}
     J_{+}^{a}(z)J_{+}^{b}(w) =\frac{kM^{ab}}{(z-w)^2}
     +\frac{f^{ab}_{c}J^c(w)}{z-w}+\cdot\cdot\cdot.
\ee
\par {In a traditional ansatz energy-momentum tensor could be constructed 
from normal ordering of two currents:}
\be 
     T(z) = L^{ab}:J_aJ_b:(z).
\ee
\par {In order to have a conformal field theory energy-momentum tensor and 
currents have to satisfy the following OPE:  
\be \label{tj}
T(z)J^a(w) = \frac{J^a(w)}{(z-w)^2}+\frac{\p J^a(w)}{(z-w)}+\cdot\cdot\cdot.
\ee
From the OPE's (\ref{jj}) and (\ref{tj}), one finds that the coefficients
$ L^{ab} $ are restricted to the solutions of the Affine-Virasoro
Master-Equation (AVME):}
\be  \label{me}
    L^{ab}=kL^{ac}M_{cd}L^{db}-L^{cd}L^{ef}f_{ce}^{a}f_{df}^{b}
    -(L^{cd}f_{ce}^{f}f_{df}^{a}L^{be}
    +(a \put(20,3){\vector(-1,0){20} \vector(1,0){20}} \put(23,0) {b)).}
\ee
\par {After constructing energy-momentum tensor with some 
solution of the AVME, one can find Virasoro's central charge from 
OPE's of the energy-momentum tensors: 
\be
   T(z)T(w)=\frac{c/2}{(z-w)^4}+\frac{2T(w)}{(z-w)^2}
	    +\frac{\D T(w)}{z-w}+\cdot\cdot\cdot.
\ee
to be:} 
\be  \label{cch}
  c = kM_{ab}L^{ab}.
\ee
\par {Now we solve the general AVME (\ref{me}) exactly by supposing that $L$ 
is a symmetric and group invariant form (i.e. equation (\ref{inv}) holds 
when replacing $M$ by $L$).
The latter assumption is legitimate because the energy-momentum tensor does not 
take any change when one changes the currents by the action of the 
group\footnote{This assumption is violated when one considers only 
the action of some subgroup in coset-constructions. }. Using this invariance 
(\ref{inv} for L) to AVME: 
\begin{center}
$      L^{ab}=kL^{ac}M_{cd}L^{db}+L^{cd}L^{ea}f_{ce}^{f}f_{df}^{b} $
 $   -(L^{cd}f_{ce}^{f}f_{df}^{a}L^{be}     $
 $   +(a \put(20,3){\vector(-1,0){20} \vector(1,0){20}} \put(23,0) {b))} $
\end{center}
\begin{center}
      $    =kL^{ac}M_{cd}L^{db}-L^{cd}f_{ce}^{f}f_{df}^{a}L^{be}  $
\end{center}
\begin{center}
      $    =kL^{ac}M_{cd}L^{db}-L^{ad}f_{ec}^{f}f_{df}^{c}L^{be},  $
\end{center}
and using (\ref{cvom}), we get:}
\be   \label{lm}
      L=(kM+c_v\om)^{-1}.
\ee
\par {This result is very pleasant. It can be considered as renormalization 
of $kM$ corresponding to renormalization of $k$ in the ordinary WZNW models. 
As a special case for semi-simple groups one can choose $M$ to be $\om$ 
then from (\ref{lm}): 
\be
L = \frac{\om^{-1}}{k+c_v}
\ee
which is just the Sugawara construction. More interesting is the $SO(3,1)$
group with $M$ defined in (\ref{mso3}): 
\be
L_{SO_M(3,1)}=\frac{1}{k+c_v}\frac{1}{1+\eta^2}
		   \pmatrix{{\bf 1}_3 & \eta{\bf 1}_3 \cr 
			    \eta{\bf 1}_3 & {\bf -1}_3 \cr}
\ee
where $\eta=\a k/(k+c_v)$,
and by (\ref{cch}) the Virasoro central charge is:} 
\be\label{ccc}                                                         
c=6\frac{k(k+c_v)+\a^2 k^2}{(k+c_v)^2+\a^2 k^2}.
\ee
\par{Notice that $k$ is an integer but there is no restriction 
on $\a$, thus the above expression indicates that we can find both
irrational and rational central charges which were impossible for 
$ SO_{\Omega}(3,1) $ models.} 
\par {The solution (\ref{lm}) also recovers the results of \cite{Kiritsis}
and is in agreement with \cite{Nouri}.
\par{The next step is to find the conformal weights of the Kac-Moody primary
fields.
First let us construct the Virasoro operator $ L_0 $ for $ SO_{M}(3,1) $:}
\be
  L_0 = L^{ab}J_{0a}J_{0b}
\ee
\be
      = L^{ab}T_{0a}T_{0b}
\ee
\be
      =\frac{k+c_v}{(k+c_v)^2+k^2 \a ^2}(J^2-K^2)
      +\frac{2k \a}{(k+c_v)^2+k^2 \a ^2} (J\cdot K).
\ee
\par{In an irreducible representation of the group, Casimirs of the group 
are just c numbers, so any state in an irreducible representation of the 
group is an eigenstate for $ L_0 $. As in \cite{Shapiro} an
irreducible representation is characterized by two parameters $ j_0 $ and
$ j_1 $, where $ j_0 $ is an integer or half integer number which denotes 
the lowest spin in this representation and $ j_1 $ is an arbitrary complex
number. In terms of these parameters the Casimirs of the group are:}
\be
J^2-K^2=1-(j_0^2 + j_1^2),
\ee
\be
J\cdot K= i j_0 j_1.
\ee
\par{In this notation, conformal weights for $(j_0,j_1)$ representations are:}
\be\label{del}
  \Delta =\frac{k+c_v}{(k+c_v)^2+k^2 \a^2} \{ 1-(j_0^2 + j_1^2) \}
      +\frac{2k \a}{(k+c_v)^2+k^2 \a ^2} (i j_0 j_1).
\ee 
\par{Let us look at unitary representations of the Lorentz group. First 
consider  the main series in which $ j_0 $ is arbitrary and $ j_1 $ is  
purely imaginary. Putting $ j_1 =is $ :}
\be
  \Delta_1 =\frac{k+c_v}{(k+c_v)^2+k^2 \a ^2}(1+s^2-j_0^2 )
      -\frac{2k \a}{(k+c_v)^2+k^2 \a ^2} ( j_0 s).
\ee
\par{Other unitary representations of the Lorentz group are supplementary
representations in which $ j_0 = 0 $ and $ j_1 $ is a real number such that
$ 0 < |j_1| \leq 1 $ , in this case conformal weights are:}
\be
  \Delta_2 =\frac{k+c_v}{(k+c_v)^2+k^2 \a ^2}(1-j_1^2 ).
\ee
\par{In the supplementary representations one can introduce the following
parameter which we call it effective $k$:
\be 
  k_{eff} = k +\frac{k^2 \a ^2}{k+c_v}.
\ee
\par{In terms of this new parameter conformal weights for supplementary
representations and central charges can be expressed as follows:}
\be
  \Delta_2 =\frac{1}{k_{eff}+c_v}(1-j_1^2 ),
\ee
\be
   c= \frac{6k_{eff}}{k_{eff}+c_v}.
\ee
\par{These are just the expressions for $ \Delta $ and $ c$ in the absence 
of $ \a $ which is the ordinary Virasoro algebra based on the 
$ SO_{\Omega}(3,1) $. It means that the conformal structures of 
$ SO_{\Omega}(3,1) $ and $ SO_{M}(3,1) $ are partially homeomorphic;
"partially" because these effective $ k_{eff} $ could be introduced only in 
supplementary representations. In the main representations one can not
find such a reparameterization to transform central charges and conformal
weights simultaneously to their values in the standard $ SO_{\Omega}(3,1) $.
This shows that these theories are generally different from standard models.} 
\par{Another aspect of the expressions in (\ref{ccc}) and (\ref{del}) is
degeneracy of central charges and conformal weights which means that for a 
large set of $k$ and $\a$ one can obtain a common central charges and
conformal weights. More explicitly, 
by transforming $(k,\a)$ to $(k',\a')$ as:} 
\be 
 k'= m k+c_v (m-1),
\ee
\be
\a'^2 = m \a^2 - m (m-1) \frac{(k+c_v)^2}{k^2}, 
\ee
for a non-zero integer parameter m, the same central charges can be 
obtained for $(k',\a')$  as for $(k,\a)$, and again for the supplementary 
representations 
the conformal weights are the same for $(k',\a')$  and $(k,\a)$ .\newline 
 {\it Non-Unitarity in Representations:}
 \par{Here we shall prove a theorem on the unitarity of representations of
 the Kac-Moody algebra based on a non-compact group.}  
\par{We shall see that unlike the case of compact base groups   
where we have a finite number of unitary representations, in general we have 
no unitary representation of the Kac-Moody algebra based on non-compact Lie 
groups.}
\par{Let $G$ be a non-compact group, and  $H$ be its maximal compact subgroup. 
Suppose Kac-Moody algebra based on the group $G$ has a highest weight unitary
representation on a vector space $V$.}
\par{The vectors in $V$ form a unitary representation $ \Lambda$ of $G$ for 
the lowest value of $ L_0$ which is infinite dimensional since
$G$ is non-compact. $V$ must also carry a unitary representation 
of  $\hat{H}$ level $k$. It is well known that unitary representations 
of  $\hat{H}$ are specified by highest weights of  $H$ belonging to:}
\be
\Gamma_H =\{\la | \la \cdot \a_H \leq k \}
\ee
where $\a_H$'s are the positive roots of  $H$ which for the moment are 
assumed to be non-zero.
\par{As it can be seen $\Gamma_H $ is a finite set. This contradicts 
the well known fact that $\Lambda$ includes an infinite set of 
inequivalent representations of  $H$. Hence the unitarity of representations  
of $\hat{G}$ and $\hat{H}$ are contradictory.}
\par{On the other hand, when $\a_{H}=0 $ there is no restriction on the
representation of  $H$ which is now an abelian subgroup.}
The above statements can be summarized in the following theorem:

{\it For a Kac-Moody algebra based on a noncompact group, with a 
non-abelian  maximal compact subgroup, highest weight unitary representations 
do not exist.}

\par{According to this theorem the commutative nature of the maximal compact
subgroup is a necessary condition for the existence of a unitary 
representation for the Kac-Moody algebra\footnote{The sufficient condition 
is not obvious. However, there are some well known examples 
such as $SU(1,1)$ and $E^c_2$ which are non compact with an abelian maximal 
compact subgroup and admit 
unitary representations \cite{Hwang,Kiritsis}.}.}

\section{Effective Action, Perturbative Results and Functional Approach:}
\setcounter{equation}{0}
\par {In the last section we had an algebraic approach to the problem.
In this section we reproduce some of our results in the previous 
section among other things using a path integral formulation. This will shed 
light on the intricacies involved in the nonstandard bilinear form.
In particular it will make clear why the master equation has the solution
(\ref{lm}).
For a path integral formulation of WZNW models, firstly consider the WZNW
action on a curved manifold  $\Sigma$ : 
\be
    S\{U|\g\} = \frac{1}{4\la^2}\int_{\Sigma}d^2x\sqrt{\g}\g^{\mu\nu}
	 <\D_{\mu}U,\D_{\nu}U^{-1}>_M+k\Gamma_{WZ}
\ee
where $\Gamma_{WZ}$ is Wess-Zumino term 
(the three dimensional integral in (\ref{ggwz1})) and 
$\g^{\mu\nu}$ and $ \g $ are the inverse and the determinant of
the metric tensor, respectively.} 
\par {For this chiral model let us look at the 
holomorphic part of the theory. To find correlation functions for the  
holomorphic currents $ J $, one can introduces sources $ l_{\mu}^a(x) $,
and add a source term to the action as:}
\be
S\{U|\g,l\}=S\{U|\g\} + \int d^2x \sqrt{\g}l^a_{\mu}(x)J^{\mu}_a(x).
\ee
\par {The effective action of the model can be introduced by the following 
path integral:}
\be
       e^{-S_{eff}(\g,l)} = \int DUe^{-S(U|\g,l)}.
\ee
\par {All the connected correlation functions for currents and energy-momentum 
tensor could be
derived by taking derivatives of $ S_{eff} $ with respect to $l$ and/or 
$\g^{\mu\nu}$.

To study quantum theory it is convenient to set: 
\be
    U(x)=U_{cl}(x)U_{q}(x),
\ee
in the action, where $U_{q}$ is quantum fluctuation around the classical 
field $U_{cl}$ which satisfies classical equations of motion. By applying   
generalization of Polyakov-Wiegmann equation \cite{Polyakov}, we obtain:
\be
   S\{U|\g,l\} = S\{U_{cl}|\g,l\} + S\{U_q|\g,0\} + 2i \int 
   d^{2}x{\g}\g^{\mu\nu}<R_{\mu},\p_{\nu}U_qU^{-1}_q>_M
\ee
where $R_{\mu}$ is the classical antiholomorphic current which equals to 
zero by using the equation of motion \cite{Leut}. Therefore, we can write: 
\be 
   S_{eff}\{\g,l\} = S_{cl}\{\g,l\} + S_q\{\g\},
\ee
in which $S_{cl}$ depends on both metric and sources explicitly, 
and $S_q\{\g\}$  is quantum effective action defined by: 
\be 
   e^{S_q\{\g\}} = \int DU_q e^{S_q\{U|\g\}},
\ee
and does not depend on sources at all, so all the correlation functions 
can be obtained from the classical action.
This feature has a crucial role in solubility of the \w models which 
also exists in our extended case.

Despite this fact we still need to calculate $ S_q $ since it gives the
energy momentum tensor and conformal anomaly.

Since current-current correlation functions could be found by taking 
derivatives of $S_{eff}$ with respect to $l_{\mu}(x)$, the quantum part 
$S_q$ does not contribute to these correlation functions, i.e. all loop 
corrections will cancel out \cite{Leut}. Firstly, let us   
calculate these current-current correlation functions when the external 
sources vanish:
\be
   <J^a_{\mu}(x)J^b_{\nu}(y)>_{conn.}= -2\pi k M^{ab}\p^x_{\mu}
       \p^y_{\nu}G(x,y).
\ee
Note that this is a connected correlation function.
\par {Turning on the external sources we find the correlation function: 
\be 
  <J^a_{\mu}(x)J^b_{\nu}(y)>_l=\frac{\del S_{eff}}{\del l^{\mu}_a(x)} 
  \frac{\del S_{eff}}{\del l^{\nu}_b(y)} 
  - \frac{\del^2 S_{eff}}{\del l^{\mu}_a(x) \del l^{\nu}_b(y)}.
\ee
It consists of disconnected (first term) and connected (second term) 
parts of the correlation function. After some calculation one finds:}
\be
     <J^a_{+}(x)J_{+a}(y)>_{disc.}=2 k^2
       <\D_{+}U_{cl}(y),\D_{+}U_{cl}^{-1}(y)>_M,  
\ee
\be
 <J^a_{+}(x)J_{+a}(y)>_{conn.}=-2\pi k M^{ab}M_{ab}\D_{+}^x 
    \D_{+}^y G(x,y)+2kc_v <\D_{+}U,\D_{+}U^{-1}>_{\om}.
\ee
\par {The first term in the second equation is just the connected 
contribution in the absence of the external fields, and the last term  
renormalizes the disconnected part by renormalizing $ kM $ in the 
disconnected part to $ kM + c_v\om $. 
It means that turning on the external field only renormalizes the
disconnected part.
This renormalization is a reflection of the Sugawara construction 
and the solubility of the model.}
\par {In order to derive the $\beta$-function and the energy-momentum 
tensor, we try to find quantum corrections via the quantum generating 
functional $S_q$. We put $ U_q = exp(i\la\pi) $ in which 
$ \pi = \pi^a(x)T_a$ and expand it around unity:
\be    
     U_q = 1+i\la\pi-\frac{\la^2}{2}\pi^2
	     -i\frac{\la^3}{6}\pi^3+\cdot\cdot\cdot.  
\ee
Using  $ A^{a}_{\mu}T_{a} = U^{-1}\D_{\mu}U $ and 
the above expression in the action (\ref{gWZNW}) 
to the fourth order in $\pi$ one finds the following lagrangian density: 
\be  \label{lagr}  
     {\cal L} = -\frac{1}{2}M_{ab}\D_{\mu}\pi^{a}\D_{\mu}\pi^{b}-
     \frac{\la^2}{24}M_{ab}f^{a}_{mn}f^{b}_{kl}\D_{\mu}\pi^{m}\pi^{n}
     \D_{\mu}\pi^{k}\pi^{l}+\frac{k\la^3}{12\pi}
     \epsilon^{\mu\nu}M_{cd}f^{d}_{ab}
     \D_{\mu}\pi^{a}\D_{\nu}\pi^{b}\pi^{c}.  
\ee
The three dimensional integration on manifold $B$ is converted
to a two dimensional term (last term in (\ref{lagr})) by divergence 
theorem.}
\par {From the above lagrangian we can find Feynman rules for
perturbative calculations. These rules include two factors. The
ordinary part which includes the effects of numerical constants and
derivative operators, and a second part which consists of group
theoretic tensorial structure.}
\par {The main difference between the two \w models in (\ref{nWZNW}) and 
(\ref{gWZNW}) is 
their tensorial properties which will be manifested in the second part of
the Feynman rules. So, let us consider the group theoretic or the 
tensorial structural of the Feynman rules:} 
\begin{center}         
	 \be
	     \put(-125,40){a\line(1,0){30}b}\put(70,40){$M^{ab}$}
	     \put(-125,-
10){n\line(1,1){20}k}\put(70,0){$M_{ab}f^{a}_{mn}
	     f^{b}_{kl}$}
	     \put(-127,10){m\line(1,-1){20}l}
	     \put(-125,-60){a\line(1,1){10}}\put(70,-
50){$M_{cd}f^{d}_{ab}$}
	     \put(-125,-40){b\line(1,-1){10}}\put(-110,-
50){\line(1,0){10}c}  
	 \ee
\end{center}
\par {The $\beta$-function could be derived considering the quantum 
corrections to the two point propagator. For first order of approximation, 
consider one loop corrections in fig.1.a and b. Corresponding expressions 
for each graphs are:\footnote{Notice that for complete expressions we 
consider permutations of partial derivatives on Green functions, but 
for brevity we write here only one form of these permutations. } }
\begin{center}
$
 \Pi^{ab}_1(x_1,x_2) = (\frac{k\la^3}{12\pi})^2 c_v M^{ac} \om_{cd} M^{db} 
		      \int d^2xd^2y \sqrt{\g(x)}
	      \sqrt{\g(y)} \epsilon^{\mu\nu}\epsilon^{\rho\sigma}  $
\end{center}
\be 
 \times \p^x_{\mu}\p^y_{\rho}G(x,y)\p^x_{\nu}\p^{y\sigma}G(x,y)
     G(x_1,x)G(y,x_2),
\ee
\be
    \Pi^{ab}_2(x_1,x_2) = -\frac{\la^2}{24} c_v M^{ac} \om_{cd} M^{db} 
      \int d^2x \sqrt{\g(x)}\p^x_{\mu}G(x,x)\p^{x\mu}G(x_1,x)G(x,x_2).
\ee
From which after regularization\footnote{Regularizations of such integrals 
including two dimensional $\epsilon$ tensors are fully developed 
in \cite{DeWitt}. } the $\beta$-function could be derived as:
\be
       \beta^{ab} =-\frac{\la^2 c_v}{4\pi} M^{ac} \om_{cd} M^{db} 
	     \{ 1-(\frac{k \la^2}{2\pi})^2 \}. 
\ee
\par {Its fixed point will be at $k=\frac{2\pi}{\la^2}$ the same as 
the ordinary \w.} 
\newline
\unitlength=1.00mm
\linethickness{0.4pt}
\begin{picture}(103.00,20.00)
\put(38.00,00){\circle{8.00}}
\put(34.00,00){\line(-1,0){8}}
\put(42.00,00){\line(1,0){8}}
\put(81.00,4.00){\circle{8.00}}
\put(71.00,00){\line(1,0){20}}
\put(124.00,00){\circle{8.00}}
\end{picture}
\newline
\put(36,0){(a)} \put(79,0){(b)} \put(123,0){(c)} 
\begin{center}
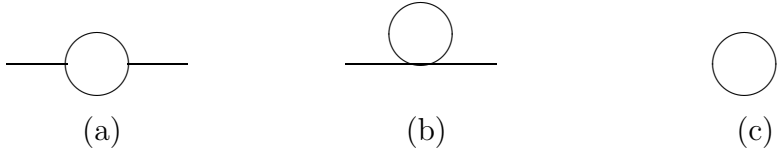
{\it Figure 1. One Loop (Propagators and Generating Functional).}
\end{center}         

\par {In order to calculate $S_q$, let us restrict ourselves to this 
fixed point. In the first order of perturbation we obtain:
\be
  e^{S_q}= \int d'\pi e^{\int d^2 \sqrt{\g}<\p_{\mu}\pi,\p^{\mu}\pi>_M}, 
\ee        
from which, it is evident that:
\be
     S^{(1)}_q\{\g\} = dim(G) D\{\g\} 
\ee     
where:
\be
    D\{\g\} = \frac{1}{2}(ln det'(-\Delta)-lnV),
\ee    
in which $\Delta$ and $V$ are the laplacian and volume of the  manifold 
$\Sigma$, respectively, and prime denotes that the zero modes of 
the laplacian are to be omitted when evaluating the determinant.
This one loop result corresponds to the graph in fig.1.c.}
\par {For two loops, we need to consider graphs in fig.2. For each 
graph we have a contribution to the generating functional:
\be
 S_a =\frac{\pi}{3}\frac{c_v}{k}Tr(M^{-1}\om)\int d^2x\g[\p^x_{\mu}G(x,y)
      \p^{y\mu}G(x,y)-G(x,y)\p^x_{\mu}\p^{y\mu}G(x,y)]_{x=y}, 
\ee
\be
 S_b =\frac{2\pi}{3}\frac{c_v}{k}Tr(M^{-1}\om)\int d^2xd^2y\g
      \epsilon^{\mu\nu}\epsilon^{\rho\sigma}
      \p^x_{\mu}\p^y_{\rho}G(x,y)\p^x_{\nu}\p^{y\sigma}G(x,y)G(x,y)
\ee
which make sense after regularization. Moreover, we ought to add a   
contribution from the path integral measure due to changing variables
from $U$ to $\pi$. The measure contribution $S_m$ to the quantum effective
action has been calculated in the appendix to be:
\be
  S_m =\frac{\pi}{3}\frac{c_v}{k}Tr(M^{-1}\om)\int d^2x\g G(x,y)[\del(x,x)+ 
       \frac{2}{V}].
\ee
As stated in the appendix, the path integral measure can be constructed
by any non-degenerate bilinear form $\Xi$, however, $S_m$ and hence $S_q$ 
and its derivatives do not depend on the choice of $\Xi$.}
\newline
\unitlength=1.00mm
\linethickness{0.4pt}
\begin{picture}(51.00,23.00)
\put(67.00,8.00){\circle{8.00}}
\put(67.00,00){\circle{8.00}}
\put(95.00,5.00){\circle{14.00}}
\put(88.00,5.00){\line(1,0){14}}
\end{picture}
\newline
\put(65,0){(a)}   \put(93,0){(b)}  
\begin{center}
{\it Figure 2. Two Loops.}
\end{center}         

\par {To regularize the above expressions for $S_a$, $S_b$ and $S_m$ 
we use the method of \cite{Leut} with new bilinear form $M$. Putting all 
together we find the generating functional up to two loops order:}
\be \label{sq}
  S^{(2)}_q =S^{(1)}_q+S_a+S_b+S_m
\ee
\begin{center}
  $ =D\{\g\} Tr[(M^{-1}-\frac{c_v}{k}M^{-1}\om)M].  $
\end{center}
\par {In higher loops, by using the three properties of $M$ 
(\ref{symm},\ref{inv},\ref{invers}), it is possible 
to transform the tensorial part of each graph at each loop order into a 
unique form. This means that the choice of different bilinear form consistent 
with (\ref{symm},\ref{inv},\ref{invers}) and addition of the tensorial part 
to the Feynman rules does not effect the solubility of the model. This is 
manifested in (\ref{sq}) when we find the two loop generating functional 
to be the same as one loop result up to a multiplicative factor.} 
\newline
\unitlength=1.00mm
\linethickness{0.4pt}
\begin{picture}(91.00,55.00)
\put(16.00,15.00){\circle{14.00}}
\put(16.00,15.00){\line(0,1){7}}
\put(16.00,15.00){\line(-3,-2){6}}
\put(16.00,15.00){\line(3,-2){6}}
\put(41.00,15.00){\circle{14.00}}
\put(35.00,18.00){\line(1,0){12}}
\put(35.00,12.00){\line(1,0){12}}
\put(65.00,22.00){\circle{6.00}}
\put(65.00,16.00){\circle{6.00}}
\put(65.00,10.00){\circle{6.00}}
\put(84.00,13.00){\circle{12.00}}
\put(84.00,22.00){\circle{6.00}}
\put(78.00,13.00){\line(1,0){12}}
\put(13.00,42.00){\circle{12.00}}
\put(19.00,42.00){\circle{12.00}}
\put(41.00,42.00){\circle{14.00}}
\put(41.00,49.00){\line(-1,-2){6}}
\put(41.00,49.00){\line(1,-2){6}}
\put(84.00,42.00){\circle{14.00}}
\put(84.00,52.00){\circle{6.00}}
\put(84.00,49.00){\line(0,-1){14}}
\end{picture}
\begin{picture}(30.00,26.00)
\put(23.00,19.00){\line(1,0){20}}
\put(28.00,26.00){\line(2,-3){10}}
\put(38.00,26.00){\line(-2,-3){10}}
\put(28.00,26.00){\line(1,0){10}}
\put(23.00,19.00){\line(2,-3){5}}
\put(43.00,19.00){\line(-2,-3){5}}
\end{picture}
\begin{center}         
{\it Figure 3. Three Loops.}
\end{center}         

\par {Let us look at the three loops as an illustration. There are eight
3-loops graphs as in fig.3. One can easily check by using the properties
of $M$ and the Jacobi identity for structural constants
that all of these graphs could be converted to a unique tensorial form as 
follows:}
\be
  (\frac{c_v}{k})^2 Tr[M^{-1}\om M^{-1}\om M].
\ee
\par {In an obvious manner, we generalize this expression to N-loop:  
\be
  (\frac{c_v}{k})^{N-1} Tr[(M^{-1}\om)^{N-1} M].
\ee
Taking account of the numerical coefficients,
this is manifestly the trace of Nth-term in the expansion of 
$(kM+c_v\om)^{-1}$ times $M$, so: 
\be
  S_q = D\{\g\}Tr[k(kM+c_v\om)^{-1} M]. 
\ee
In each order for $S_q$, the results of \cite{Leut} could be 
recovered by replacing $M$ by $\om$ in $S_q$ for a semi-simple group.}   
By using (\ref{cch}) we obtain:
\be
  S_q = c D\{\g\}.
\ee
\par {Thus, in this category of \w, we have found the same expression for  
$S_q$ as the ordinary \w models in terms of the central charge $c$ and 
$D\{\g\}$, from which we find the trace anomaly to all orders of 
perturbation to be \cite{Leut}:
\be
T^{\mu}_{\mu} =\frac{c}{24\pi}R,
\ee
where $R$ is the two dimensional scalar curvature. This result is 
the same as that the ordinary \w models based on semi-simple groups.} 
\par {The renormalization of $kM$ can also be obtained by the functional 
method. Consider the partition function of the model:}
\be
  Z =\int dU e^{kS_M\{U|\g\}}. 
\ee
\par {Change the variables of integration from $U$ to $A$:  
\be \label{ua}
  A^a_{+}T_a=U^{-1}\p_{+}U
\ee   
which introduces a Jacobian in the path integral measure:}
\be
  dU=dA det(\frac{\p U}{\p A_{+}}).
\ee
From (\ref{ua}) the determinant can be found to be \cite{Polyakov2}:
\be
det(\frac{\p U}{\p A_{+}})=det[\Delta_{+}(A)]
\ee
where $ \Delta_{+}(A) =\p_{+}+[A_{+},\  ]$ and can be calculated 
using ghost fields integration technique:
\be \label{det}
det[\Delta_{+}(A)] = \int db dc e^{<b,\Delta_{+}c>_{\Xi}}
\ee
\be
    = \int db dc e^{<b,\p_{+}c>_{\Xi} + <b,[A_{+},c]>_{\Xi}}.
\ee
We shall apply bosonization techniques \cite{Polyakov2} to 
calculate the above determinant. 
\par {From the $bc$ fields one can find corresponding 
conserved current to be:
\be
J^{(gh)}_{+j}=f^i_{mn}b^n c^m {\Xi}_{ij}.
\ee
These currents will obey the Kac-Moody algebra with $c_v$ as the central 
charge by the following OPE:
\be \label{jgh}
     J_{+a}^{(gh)}(z)J_{+b}^{(gh)}(w) =\frac{c_v\om_{ab}}{(z-w)^2}
     +\frac{f_{ab}^{c}J_{+c}^{(gh)}(w)}{z-w}+\cdot\cdot\cdot.
\ee
So, the effective action for the ghost fields will be the \w action for 
gauge fields $A$, i.e.:
\be
det[\Delta_{+}(A)] = e^{c_vS_{\om}\{A|\g\}}
\ee
where,
\be
    S_{\om}\{A|\g\} = \int d^2x\g\g^{\mu\nu}
	 <A_{\mu},A_{\nu}>_{\om}+\Gamma_{WZ}
\ee
whose current algebra is just (\ref{jgh}). It is worth noting that as long 
as $\Xi$ is nondegenerate, (\ref{det}) is independent of $\Xi$.}
\par {Putting these results into the path integral one find the following 
partition function:
\be 
Z=\int dA e^{kS_M(A) + c_vS_{\om}(A)}.
\ee
It manifestly demonstrates the $kM$ renormalization to $kM+c_v\om$.}
\section{Conclusions}
\par {In this work WZNW models for noncompact and non-semisimple lie groups
are developed. Firstly for  
non-compact groups we proved an important theorem that the highest weight
representations of Kac-Moody algebras for these groups are nonunitary
when maximal compact subgroup of the base group is nonabelian. On the other 
hand, in such cases the third homotopy group of the lie group is nontrivial  
which leads to integer Kac-Moody central charges $k$. However, using a  
nonstandard bilinear form which couples compact and noncompact subspaces as
in the case of $SO(3,1)$, we obtained irrational conformal weights and 
central charges.}  
\par {Hence, introducing nonstandard bilinear forms is useful in developing 
new conformal theories in the case of non-compact groups and is necessary 
in the case of non-semisimple groups. In this relation we treated the model
perturbatively and showed that $\beta$-function vanishes at the same points 
as in standard models.} 
\par {Renormalization of $kM$ was investigated in different approaches. 
Firstly, it is due to the renormalization of disconnected graphs in 
current-current correlations. Secondly, by calculating quantum generating 
functional of two dimensional metric to all orders of perturbation we found 
the renormalized $kM$ in the Sugawara form. Finally, this renormalization 
can be seen from functional determinant. From the generating functional 
it is easy to obtain the trace anomaly.}
\par {These approaches show that the nonstandard WZNW theories can serve as 
novel conformal field theories. There are still open questions such as 
existence of unitary representations of the Virasoro algebra which shall be 
addressed elsewhere.} 
\newline
{\large {\bf Acknowledgment}}
\par{We would like to thank F. Ardalan, A. Ghezelbash, V. Karimipour, 
A. Mostafazadeh, M. Rahimi Tabar and A. Shariati for useful discussions.} 
\section{Appendix}
\setcounter{equation}{0}
\par {Here, we treat the path integral measure in terms of $\pi$ fields.
To define a measure we need to define distance notion in field 
configuration space. It can be done by using some bilinear form $\Xi$ as 
follows:}
\be
d^2=\int d^2x \sqrt{\g} <dU(x),dU^{-1}(x)>_{\Xi}.
\ee
\par {Now decompose $U$ as $ U=U_0 U'$, where $ U_0 $ and $ U' $ describe 
zero modes and nonzero modes of the Laplace operator, respectively. 
Then the distance becomes:
\be
d^2=\int d^2x \sqrt{\g}\{ <dU_0,dU^{-1}_0>_{\Xi}-2<U_0^{-1}dU_0,
		      dU'U'^{-1}>_{\Xi} + <dU',dU'^{-1}>_{\Xi} \}.
\ee
We can arrange it as follows: 
\be
d^2=V\int d^2x \sqrt{\g} <(iU_0^{-1}dU_0+u)^2>_{\Xi} + d'^2,
\ee
where:}
\be \label{d'2}
d'^2=\int d^2x \sqrt{\g} <dU',dU'^{-1}>_{\Xi} - <u^2>_{\Xi},
\ee
\be
u=\frac{i}{V} \int d^2x \sqrt{\g} dU'U'^{-1}.
\ee
\par {Since $d'^2$ is independent of collective variable $U_0$, the 
measure factorize to:}
\be
dU= V^{1/2dimG} d\mu (U_0)[dU'].
\ee
\par {Now putting $U'=exp(i\la \pi)$  and expand in powers of $\la$ one  
obtains:}
\be   \label{du'}
 <dU',dU'^{-1}>_{\Xi}=\la^2<d\pi,d\pi>_{\Xi} 
	     +\frac{\la^2}{12}<[d\pi,\pi]^2>_{\Xi}+\cdot\cdot\cdot,
\ee
\be   \label{u}
u=\frac{i\la^2}{2V} \int d^2x \sqrt{\g} [d\pi,\pi] +\cdot\cdot\cdot.
\ee
\par {$\pi$ fields could be parameterized in terms of hermitian matrix 
functions $g_r$ which span the space of none zero modes as follows:
\be
\pi (x)=\Sigma_r g_r(x) q^r.
\ee
Then the distance $d'^2$ will be:
\be   \label{grs}
d'^2=\Sigma G_{rs} (q)dq^r dq^s.
\ee
From this expression the volume element can be read as:}
\be 
dU'=(det\  G)^{1/2} \Pi_r dq^r.
\ee
\par {Expanding the metric $G$ in powers of $\la$, we find:
\be
G_{rs}=\la^2 G_{rs}^{(0)}+\la^4 G_{rs}^{(1)}(q) 
       +\la^6 G_{rs}^{(2)}(q)+\cdot\cdot\cdot,
\ee
in which $G_{rs}^{(0)}$ does not depend on the variables $q$. From the 
last equation $detG $ can be determined as follows: 
\be
(detG)^{1/2} \sim (detG^{(0)})^{1/2}(1+\la^2 m_1 +\la^4 m_2 +O(\la^6))
\ee
where:}
\be
m_1=\frac{1}{2}Tr(G^{(0)^{-1}}G^{(1)}),
\ee
\be
m_2=\frac{1}{2}Tr(G^{(0)^{-1}}G^{(2)})+\frac{1}{8}[Tr(G^{(0)^{-1}}G^{(1)})]^2
	-\frac{1}{4}Tr(G^{(0)^{-1}}G^{(1)})^2.
\ee
\par {By substituting (\ref{du'})  and (\ref{u}) in (\ref{d'2}) for $d'^2$,
$G$ can be read from (\ref{grs}): 
\begin{center}
$ G_{rs}=\la^2 \int d^2x \sqrt{\g} <g_r,g_s>_{\Xi} 
	+\frac{\la^4}{12}\int d^2x \sqrt{\g} <[\pi,g_r],[\pi,g_s]>_{\Xi} $ 
\end{center}
\be  
  +\frac{\la^4}{4V} \int d^2x d^2y \sqrt{\g} <[\pi_x,g_r],[\pi_y,g_s]>_{\Xi}.
\ee
Using the completeness of $g_r$ matrices which means that:
\be
(G^{(0)^{-1}})^{rs} g_r(x) \otimes g_s(y) = (\delta (x,y)-\frac{1}{V}),
\ee
it can be seen that:
\be \label{m1}
m_1=\frac{1}{2}Tr(G^{(0)^{-1}}G^{(1)})=\frac{-c_v}{12} \int d^2x 
	 \sqrt{\g} <\pi,\pi>_{\om}(\delta (x,x)+\frac{2}{V}).
\ee
Contribution of this term to the action can be easily found to be :}
\be
S_{m_1}=\frac{c_v}{6k}Tr(M^{-1}\om) \int d^2x \sqrt{\g} G(x,x) 
		       (\delta (x,x)+\frac{2}{V}) .
\ee
\par {It is worth mentioning that from (\ref{m1}) $m_1$ is independent of 
$\Xi$  and this is also true for $m_2$ and higher order terms.}

\end{document}